\documentclass[a4paper]{article}

\usepackage{INTERSPEECH2021}

\usepackage[nolist,nohyperlinks]{acronym}
\usepackage{comment}
\usepackage{hhline}

\usepackage{pifont}
\newcommand{\cmark}{\ding{51}}%
\newcommand{\xmark}{\ding{55}}%
\usepackage{amssymb}
\usepackage{amsmath}
\usepackage{amsfonts}
\usepackage{url}

\acrodef{TTS}{Text-to-Speech}
\acrodef{SPSS}{Statistical Parameter Speech Synthesis}
\acrodef{MA}{Moving Average}
\acrodef{AR}{Auto-Regressive}
\acrodef{PWG}{Parallel WaveGAN}
\acrodef{QoE}{Quality of Experience}
\acrodef{BWE}{Bandwidth Expansion}
\acrodef{WGAN}{Wasserstein GAN}
\acrodef{AGD}{Alternating Gradient Descent}
\acrodef{UDA}{Unsupervised Domain Adaptation}
\acrodef{VEA}{Visual Emotion Analysis}
\acrodef{SOTA}{state-of-the-art}
\acrodef{PLDA}{Probabilistic Linear Discriminant Analysis}
\acrodef{DFD}{Deep Feature Discriminator}
\acrodef{LSGAN}{Least Squares GAN}
\acrodef{PCA}{Principal Component Analysis}
\acrodef{CNN}{Convolutional Neural Network}
\acrodef{TF}{Time-Frequency}
\acrodef{SRE}{Speaker Recognition Evaluation}
\acrodef{CTS}{Conversational Telephone Speech}
\acrodef{VAD}{Voice Activity Detection}
\acrodef{LMFB}{Log Mel-Filter Bank}
\acrodef{FFT}{Fast Fourier Transform}
\acrodef{LR}{Learning Rate}
\acrodef{IN}{Instance Normalization}
\acrodef{GLU}{Gated Linear Units}
\acrodef{MVN}{Mean-Variance Normalization}
\acrodef{SNR}{Signal-to-Noise Ratio}
\acrodef{EER}{Equal Error Rate}
\acrodef{minDCF}{Minimum Decision Cost Function}
\acrodef{LDA}{Linear Discriminant Analysis}
\acrodef{t-SNE}{t-distributed Stochastic Neighbor Embedding}
\acrodef{UMAP}{Uniform Manifold Approximation and Projection}
\acrodef{SGD}{Stochastic Gradient Descent}
\acrodef{NIST}{National Institute of Standards and Technology}

\title{Deep Feature CycleGANs: Speaker Identity Preserving Non-parallel Microphone-Telephone Domain Adaptation for Speaker Verification}
\name{Saurabh Kataria, Jes\'us Villalba, Piotr \.Zelasko, Laureano Moro-Vel\'azquez, Najim Dehak}
\address{Center for Language and Speech Processing,
  Human Language Technology Center of Excellence, \\
    Johns Hopkins University, Baltimore, MD, USA}
\email{\{skatari1, jvillal7, pzelasko, laureano, ndehak3\}@jhu.edu}

\begin{document}

\maketitle
\begin{abstract}
With the increase in the availability of speech from varied domains, it is imperative to use such out-of-domain data to improve existing speech systems.
Domain adaptation is a prominent pre-processing approach for this.
We investigate it for adapt microphone speech to the telephone domain.
Specifically, we explore CycleGAN-based unpaired translation of microphone data to improve the x-vector/speaker embedding network for Telephony Speaker Verification.
We first demonstrate the efficacy of this on real challenging data and then, to improve further, we modify the CycleGAN formulation to make the adaptation \emph{task-specific}.
We modify CycleGAN's identity loss, cycle-consistency loss, and adversarial loss to operate in the \emph{deep feature} space.
\emph{Deep features} of a signal are extracted from an auxiliary (speaker embedding) network and, hence, preserves speaker identity.
Our 3D convolution-based Deep Feature Discriminators (DFD) show relative improvements of 5-10\% in terms of equal error rate.
To dive deeper, we study a challenging scenario of pooling (adapted) microphone and telephone data with data augmentations and telephone codecs.
Finally, we highlight the sensitivity of CycleGAN hyper-parameters and introduce a parameter called \emph{probability of adaptation}.
\end{abstract}

\section{Introduction}
\label{sec:intro}
Telephony speaker verification draws significant interest from the speaker recognition community, thanks to the \ac{SRE} that is conducted regularly by \ac{NIST}.
The state-of-the-art system uses x-vector speaker embedding front-end~\cite{snyder2018x} and \ac{PLDA} based back-end \cite{villalba2020advances,villalba2019jhu}.
In this work, we experiment with evaluation data acquired from the telephone domain.
A standard way of constructing the training data for the x-vector network is to pool together telephone and microphone data.
However, the microphone data is downsampled (from 16 kHz to 8 kHz) and treated with standard telephone codecs like GSM, AMR-NB to simulate telephone domain~\cite{bittner2016pysox,sivaraman2020speech}.
Speech signal from a particular \emph{domain} comprises of characteristic channel and transmission effects.
We argue that a more principled approach like \emph{domain adaptation} is suited better since the telephone domain consists of other channel effects in addition to different codecs ~\cite{koster2018multidimensional}.

\ac{UDA} is suited for this scenario since access to paired data from multiple domains is usually infeasible.
CycleGAN (Cycle-Consistent Generative Adversarial Network)~\cite{zhu2017unpaired} (Sec.~\ref{sec:cyclegan}) is a popular GAN-based unpaired translation/style transfer technique.
Adaptation of speech features using CycleGAN works well for problems like voice conversion~\cite{kaneko2017parallel}, speech enhancement~\cite{nidadavolu2020unsupervised,nidadavolu2020single}, and domain adaptation~\cite{nidadavolu2019cycle,nidadavolu2019low}.
In \cite{nidadavolu2019cycle}, authors pursued microphone speaker verification by adapting microphone evaluation data to telephone domain to overcome the \emph{domain shift} w.r.t. their telephone domain training data.
In the follow-up study \cite{nidadavolu2019low}, the authors focused on the scenario where the availability of the target domain data is restricted.
Various modifications to CycleGAN are proposed in the literature, like having multiple discriminators~\cite{hosseini2018multi} and preserving class labels in the forward and backward cycle (semantic consistency)~\cite{zhao2020emotional}.
It is known that the cycle-consistency constraint is too restrictive and limits the flexibility in predictions \cite{zhao2020unpaired,park2019unsupervised,hosseini2018augmented}.
Moreover, in GANs, the loss in semantic information is compensated by matching high-level \emph{deep features} of signals via \emph{content loss}~\cite{chen2018cartoongan,kumar2019melgan}.

\emph{Deep features} of a signal refer to its activations obtained from a pre-trained auxiliary network~\cite{germain2018speech,kataria2020perceptual,zhang2020perceptual}.
Choosing it as a speaker embedding network in the \emph{content loss} minimizes the distance between the speaker embeddings of signals.
It also makes the formulation \emph{task-specific} \cite{kataria2020feature,kataria2020analysis}, since our downstream task is speaker verification.
This idea of preserving certain attributes is explored in non-CycleGAN solutions for linguistic~\cite{luong2020nautilus} and speaker identity preservation~\cite{du2020optimizing}.
The study \cite{you2021axial} defined the cycle-consistency loss in the \emph{deep feature space} for preserving phonetic information for voice conversion task.
Our goal is to devise a \emph{complete} re-formulation of CycleGAN where \emph{all} loss terms and constraints are computed in the \emph{deep feature space}.
This is similar to the \emph{task-specific} speech enhancement of \cite{kataria2020feature}.
To improve the generators and the discriminators of CycleGAN, we modify its identity, cycle-consistency, and adversarial loss.
The focus is on improving the utility of microphone training data (via adaptation) for telephony speaker verification.
Previous works have only explored adapting test sets with no speaker identity preservation \cite{nidadavolu2020unsupervised,nidadavolu2020single}.
Note that, we refer to \emph{microphone speech} as recordings that are originally wide-band, in consistency with NIST terminology.

Our contributions are: 1) using CycleGANs, we provide the first study for combining telephone and (adapted) microphone training data for Telephony Speaker Verification; 2)  we re-formulate all components of CycleGAN to operate in deep feature space, thereby alleviating cycle-consistency strictness issue and making the formulation \emph{task-specific} w.r.t. the downstream task; 3) introduce \ac{DFD} which uses deep features for real/fake determination in adversarial loss; 4) demonstrate the efficacy of proposal on three real test sets while quantifying the effect of telephone codecs, data augmentation, \emph{probability of adaptation}, and hyper-parameters.



\section{Description of CycleGAN}
\label{sec:cyclegan}
CycleGAN~\cite{zhu2017unpaired} is an unsupervised generative model which was proposed in computer vision for unpaired translation (or domain adaptation).
It is based on Generative Adversarial Networks (GANs)~\cite{goodfellow2014generative} which are briefly summarized as a minimax game between generator $(\mathcal{G})$ and discriminator $(\mathcal{D})$ network where discriminator tries to distinguish between real and fake (synthesized by $\mathcal{G}$) sample, while generator tries to fool the discriminator into believing that the synthesized sample is real.
The goal of GAN is to learn to generate real samples via $\mathcal{G}$.
CycleGAN consists of two GANs each trying to learn \emph{realistic} mapping (enforced by adversarial loss) from one domain to the other.
The training of two GANs is linked via a cycle-consistency constraint which promotes the mappings to be reversible which, desirably, restricts the output space of generators.
If the two domains microphone and telephone are denoted by $\mathcal{M}$ and $\mathcal{T}$ respectively, mathematically, we optimize:
\begin{equation}
    \label{eq:cg}
    \begin{split}
        \min_{\mathcal{G}_{\mathcal{M}\rightarrow\mathcal{T}},\mathcal{G}_{\mathcal{T}\rightarrow\mathcal{M}}} \max_{\mathcal{D}_{\mathcal{M}},\mathcal{D}_{\mathcal{T}}} \mathcal{L}_{\text{cyc-GAN}},
    \end{split}
\end{equation}
where $\mathcal{L}_{\text{cyc-GAN}}$ is given by a weighted sum of adversarial, cycle-consistency and identity losses,
\begin{equation}
\label{eq:cg2}
    \begin{split}
    &\mathcal{L}_{\text{cyc-GAN}}(\mathcal{G}_{\mathcal{M}\rightarrow\mathcal{T}},\mathcal{G}_{\mathcal{T}\rightarrow\mathcal{M}},\mathcal{D}_{\mathcal{M}},\mathcal{D}_{\mathcal{T}}) =\\
    &\enspace\mathcal{L}_{\text{GAN},\mathcal{M}}(\mathcal{G}_{\mathcal{T}\rightarrow\mathcal{M}},\mathcal{D}_{\mathcal{M}}) + \mathcal{L}_{\text{GAN},\mathcal{T}}(\mathcal{G}_{\mathcal{M}\rightarrow\mathcal{T}},\mathcal{D}_{\mathcal{T}})\\
    &+\lambda_{\text{cyc}}\mathcal{L}_{\text{cyc}}(\mathcal{G}_{\mathcal{M}\rightarrow\mathcal{T}},\mathcal{G}_{\mathcal{T}\rightarrow\mathcal{M}}) +
     \lambda_{\text{id}}\mathcal{L}_{\text{id}}(\mathcal{G}_{\mathcal{M}\rightarrow\mathcal{T}},\mathcal{G}_{\mathcal{T}\rightarrow\mathcal{M}})
    \end{split}
\end{equation}
To define the terms in~\eqref{eq:cg2}, we introduce functions $(h_{\text{cyc}}^{\mathcal{M}},h_{\text{cyc}}^{\mathcal{T}},h_{\text{id}}^{\mathcal{M}},h_{\text{id}}^{\mathcal{T}},h_{\text{disc}}^{\mathcal{M}},h_{\text{disc}}^{\mathcal{T}})$, 
which, for the vanilla Cycle-GAN in this section, are equal to identity ($\mathrm{x}=h_a^b(\mathrm{x})$). Their utility will be revealed in the next section.
$\lambda_{\text{cyc}}$ and $\lambda_{\text{id}}$ are the weights for the cycle-consistency and identity loss constraints respectively.
Identity loss constraint is a commonly used regularizer in CycleGANs generator to 1) be capable of identity functionality and 2) be conservative when presented unexpected (surprise) input.
Let $\mathbf{m}$ and $\mathbf{t}$ denote the 2D \ac{TF} features from microphone and telephone domain respectively.
Also, let $p_{\mathcal{M}}$ and $p_{\mathcal{T}}$ denote the real microphone and telephone training database respectively.
The cycle-consistency loss constraint is
\begin{equation}
    \label{eq:cyc}
    \begin{split}
        &\mathcal{L}_{\text{cyc}}(\mathcal{G}_{\mathcal{M}\rightarrow\mathcal{T}},\mathcal{G}_{\mathcal{T}\rightarrow\mathcal{M}},h_{\text{cyc}}^{\mathcal{M}},h_{\text{cyc}}^{\mathcal{T}}) = \\
        &\enspace\mathbb{E}_{\mathbf{m}\sim p_{\mathcal{M}}}[||h_{\text{cyc}}^{\mathcal{M}}(\mathbf{m}) - \mathcal{G}_{\mathcal{T}\rightarrow\mathcal{M}}(\mathcal{G}_{\mathcal{M}\rightarrow\mathcal{T}}(h_{\text{cyc}}^{\mathcal{M}}(\mathbf{m})))||_1]\\
        &\enspace+\mathbb{E}_{\mathbf{t}\sim p_{\mathcal{T}}}[||h_{\text{cyc}}^{\mathcal{T}}(\mathbf{t}) - \mathcal{G}_{\mathcal{M}\rightarrow\mathcal{T}}(\mathcal{G}_{\mathcal{T}\rightarrow\mathcal{M}}(h_{\text{cyc}}^{\mathcal{T}}(\mathbf{t})))||_1],
    \end{split}
\end{equation}
the identity loss constraint is
\begin{equation}
    \label{eq:id}
    \begin{split}
        &\mathcal{L}_{\text{id}}(\mathcal{G}_{\mathcal{M}\rightarrow\mathcal{T}},\mathcal{G}_{\mathcal{T}\rightarrow\mathcal{M}},h_{\text{id}}^{\mathcal{M}},h_{\text{id}}^{\mathcal{T}}) =\\
        &\enspace\mathbb{E}_{\mathbf{m}\sim p_{\mathcal{M}}}[||h_{\text{id}}^{\mathcal{M}}(\mathbf{m}) - \mathcal{G}_{\mathcal{T}\rightarrow\mathcal{M}}(h_{\text{id}}^{\mathcal{M}}(\mathbf{m}))||_1] \\
        &\enspace+\mathbb{E}_{\mathbf{t}\sim p_{\mathcal{T}}}[||h_{\text{id}}^{\mathcal{T}}(\mathbf{t}) - \mathcal{G}_{\mathcal{M}\rightarrow\mathcal{T}}(h_{\text{id}}^{\mathcal{T}}(\mathbf{t}))||_1],
    \end{split}
\end{equation}
and the \ac{LSGAN}~\cite{mao2017least} based (two) adversarial loss terms are
\begin{equation}
    \label{eq:gan}
    \begin{split}
        &\mathcal{L}_{\text{GAN},\mathcal{M}}(\mathcal{G}_{\mathcal{T}\rightarrow\mathcal{M}},\mathcal{D}_{\mathcal{M}},h_{\text{disc}}^{\mathcal{M}},h_{\text{disc}}^{\mathcal{T}}) =\\
        &\enspace \mathbb{E}_{\mathbf{m}\sim p_{\mathcal{M}}}[(1-\mathcal{D}_{\mathcal{M}}(h_{\text{disc}}^{\mathcal{M}}(\mathbf{m})))^2]\\
        &\enspace+ \mathbb{E}_{\mathbf{t}\sim p_{\mathcal{T}}}[(\mathcal{D}_{\mathcal{M}}(\mathcal{G}_{\mathcal{T}\rightarrow\mathcal{M}}(h_{\text{disc}}^{\mathcal{T}}(\mathbf{t}))))^2]
        ,\\
        &\mathcal{L}_{\text{GAN},\mathcal{T}}(\mathcal{G}_{\mathcal{M}\rightarrow\mathcal{T}},\mathcal{D}_{\mathcal{T}},h_{\text{disc}}^{\mathcal{M}},h_{\text{disc}}^{\mathcal{T}}) =\\
        &\enspace\mathbb{E}_{\mathbf{t}\sim p_{\mathcal{T}}}[(1-\mathcal{D}_{\mathcal{T}}(h_{\text{disc}}^{\mathcal{T}}(\mathbf{t})))^2]\\
        &\enspace+ \mathbb{E}_{\mathbf{m}\sim p_{\mathcal{M}}}[(\mathcal{D}_{\mathcal{T}}(\mathcal{G}_{\mathcal{M}\rightarrow\mathcal{T}}(h_{\text{disc}}^{\mathcal{M}}(\mathbf{m})))^2]
        .
    \end{split}
\end{equation}
$\mathcal{G}_{\mathcal{M}\rightarrow\mathcal{T}}$ and $\mathcal{G}_{\mathcal{T}\rightarrow\mathcal{M}}$ denote the \emph{forward} and \emph{backward} cycle of CycleGAN.
We employ LSGAN because it alleviates the vanishing gradient problem of the vanilla GAN \cite{mao2017least}.
To optimize Eq.~\ref{eq:cg}, we use \ac{AGD} algorithm~\cite{goodfellow2014generative} in which discriminators and generators are updated for $n_{\mathcal{D}}$ and $n_{\mathcal{G}}$ steps alternatively.

\section{Deep feature CycleGAN}
\label{sec:dfcyclegan}
As alluded in Sec.~\ref{sec:intro}, using \emph{deep features} is the core ingredient of our proposal.
Given a pre-trained auxiliary network $\mathcal{A}$, with $\mathcal{A}_{[:i]}$ denoting the network till $i^{\text{th}}$ layer, $\mathcal{A}_{[:i]}$ returns the hidden activations at layer $i$, and the corresponding deep features for input $\mathbf{x}$ using the first $q$ layers is given by
\begin{equation}
    \label{eq:dfe}
    \begin{split}
    &{\tt{DF}}(\mathbf{x},\mathcal{A},q) =\\
    &\enspace{\tt{Concat}}({\tt{Expand}}(\{\mathcal{A}_{[:1]} (\mathbf{x}), \mathcal{A}_{[:2]} (\mathbf{x}), \dots, \mathcal{A}_{[:q]} (\mathbf{x})\}))
    \end{split}
\end{equation}
where the operation ${\tt{Expand}}$ means expanding each dimension of matrices to the corresponding maximum dimension seen in the set.
The newly created entries in matrices are filled with zeros.
${\tt{Concat}}$ operation concatenates the matrices along a newly created dimension.
These operations are done to convert a set of different shape matrices to a single rectangular matrix.
For our task, the output of ${\tt{DF}}(\mathbf{x},\mathcal{A},q)$ is five-dimensional: (batch size, q, channels, height, width), which makes it suitable for processing with \ac{CNN} based architectures (Sec.~\ref{sec:exp}).

Let $\mathcal{A}_T$ and $\mathcal{A}_M$ denote the auxiliary (speaker classification) networks pre-trained with the original (unadapted) telephone and microphone training data respectively.
\emph{DF-CycleGAN} uses the following choice of $h$ functions in Eqs.~\ref{eq:cyc}-\ref{eq:gan}:
\begin{equation}
    \label{eq:dfcg}
    \begin{split}
        h_{\text{cyc}}^{\mathcal{M}} := {\tt{DF}}(\cdot,\mathcal{A}_M,q_{\text{cyc}}),&\;
        h_{\text{cyc}}^{\mathcal{T}} := {\tt{DF}}(\cdot,\mathcal{A}_T,q_{\text{cyc}})\\
        h_{\text{id}}^{\mathcal{M}} := {\tt{DF}}(\cdot,\mathcal{A}_M,q_{\text{id}}),&\;
        h_{\text{id}}^{\mathcal{T}} := {\tt{DF}}(\cdot,\mathcal{A}_T,q_{\text{id}})\\
        h_{\text{disc}}^{\mathcal{M}} := {\tt{DF}}(\cdot,\mathcal{A}_M,q_{\text{disc}}),&\;
        h_{\text{disc}}^{\mathcal{T}} := {\tt{DF}}(\cdot,\mathcal{A}_T,q_{\text{disc}})
    \end{split}
\end{equation}
$q_{\text{cyc}}$, $q_{\text{id}}$, and $q_{\text{disc}}$ are the number of layers used for deep feature extraction for cycle-consistency loss, identity loss, and adversarial loss respectively.
When modifying only one out of these three terms, we refer to the corresponding CycleGANs as \emph{DF-Cycle}, \emph{DF-Identity}, and \emph{DF-Disc.} respectively.
Note from Eqs.~\ref{eq:cyc}-\ref{eq:gan} that ${\tt{Concat}} \circ {\tt{Expand}}$ operation requires generators and discriminators to process 5D inputs.
For the generators, due to GPU memory limitations, we fuse the last two dimensions of the input tensor and simply use 2D CNNs to process the resultant 4D data.
Inspired from \cite{cai2019multi}, for (Deep Feature) Discriminators (DFD), we \emph{retain} the complete 5D data structure and use 3D CNN layers to process it.
Note that $\mathcal{A}_T$ and $\mathcal{A}_M$ are frozen during \emph{DF-CycleGAN} training.

\section{Experimental Setup}
\label{sec:exp}
We use \emph{VoxCeleb2}~\cite{nagrani2017voxceleb} and \ac{SRE} \ac{CTS} 2012-18 English telephone sets (referred to as \emph{SRE12-18}) as the training datasets for the source (microphone) and target (telephone) domains of CycleGAN respectively.
\emph{SRE12-18} is chosen to be representative of modern telephone speech.
We apply a simple energy-based \ac{VAD} pre-processing to include only voiced frames for adaptation training.
Subsequently, \emph{VoxCeleb2} consists of 5994 speakers and 2247 hrs of speech, while \emph{SRE12-18} consists of 2895 speakers and 4917 hrs of speech.

The CycleGAN is learned on 2.5 sec long 64-D \ac{LMFB} features with 25ms \ac{FFT} window and 10ms shift.
The batch size is 8, the number of epochs is 15 (1 epoch = 100 hrs data here), the \ac{LR} for generators is 0.0002, the \ac{LR} for discriminators is 0.0001, the \ac{LR} scheduler is a linearly decaying function with a minimum \ac{LR} of 1e-7, the optimizer is Adam-based
with $(\beta_1,\beta_2)=(0.5,0.999)$.
Data augmentation is not performed during CycleGAN training and \emph{VoxCeleb2} is downsampled a priori to 8 kHz to match the telephone sampling rate.
We use the generator and discriminator architectures from the Voice Conversion work of \cite{kaneko2019cyclegan}.
The generators are 14-layer encoder-decoder style deep CNNs with \ac{IN}, residual blocks, and \ac{GLU} activations.
The discriminators are simpler 5-layer CNNs.
More details on architecture can be found in \cite{kaneko2019cyclegan}.
$n_{\mathcal{D}}$ and $n_{\mathcal{G}}$ are fixed to 1 and 2 respectively, while $\lambda_{\text{id}}$ and $\lambda_{\text{cyc}}$ are set to 5 and 10 respectively.

For \emph{deep feature} based CycleGANs, $q_{\text{cyc}}=q_{\text{id}}=5$.
A higher value of $q_{\text{disc}}$ implies longer training time and memory requirements so we choose a nominal value of $q_{\text{disc}}=2$.
Auxiliary networks $\mathcal{A}_M$ and $\mathcal{A}_T$ are (pre-)trained for speaker classification task with \emph{VoxCeleb2} and \emph{SRE12-18} respectively.
The architecture used is Light Resnet-34, which is obtained from the popular Resnet-34 architecture by reducing the number of channels in all CNN layers by half.
Refer to~\cite{villalba2020advances} for more details.
Optimization is Adam-based \ac{SGD}, the number of epochs is 90 (1 epoch here means entire dataset), the batch size is 128, the loss function is AM-Softmax~\cite{wang2018additive} (margin=0.3), the sequence length is 4 sec, the \ac{LR} is 0.05, and the \ac{LR} scheduler reduces the \ac{LR} by half when validation loss does not decrease for three epochs.
Sliding-window \ac{MVN} is done on-the-fly on the input features (64-D \ac{LMFB}) with left and right context equal to 150 frames each.
Final adaptation is performed with $\mathcal{G}_{\mathcal{M}\rightarrow\mathcal{T}}$.

The x-vector network for Speaker Verification is also chosen as a Light ResNet-34 with the x-vector dimension as 256.
Its training data is different for each experiment (Sec.~\ref{sec:results}): \emph{VoxCeleb2}, \emph{SRE-Mix-SWBD}, or their combination.
\emph{SRE-Mix-SWBD} refers to the combination of SRE CTS datasets from 2004-12, MIXER6, and Switchboard, which is a standard training set used in prior work \cite{villalba2020advances}.
For experiments with data augmentation, we corrupt speech utterances with noise files from (downsampled) MUSAN~\cite{snyder2015musan} with \ac{SNR} range of $[3,18]$ dB and corruption probability of 0.7.

For evaluation, we choose \ac{PLDA} as the backend for target/non-target decision making.
The dimension for \ac{LDA} pre-processing (on x-vectors) is 150, \ac{PLDA} dimension is 125, and its training data is the combination of train and (60\% of) dev portions of CTS SRE 2016-18.
We experiment with three challenging language-mismatched real test sets from \ac{SRE} 2016 and 2019: \emph{SRE-19-eval} (Tunisian Arabic), (rest 40\%) \emph{SRE-16-yue} (Cantonese), and (rest 40\%) \emph{SRE-16-tgl} (Tagalog).
The x-vectors are extracted from a random chunk of 10-60s of speech.
Results are reported on \ac{EER} (in \%) and \ac{minDCF} $(p_{\text{tar}}=0.05)$.
The lower the \ac{EER} and \ac{minDCF}, the better.
All systems are implemented with Pytorch.

\section{Results}
\label{sec:results}
\subsection{Baseline CycleGAN adaptation}
\label{sec:baseline}
\begin{figure}
    \centering
    \includegraphics[width=\linewidth]{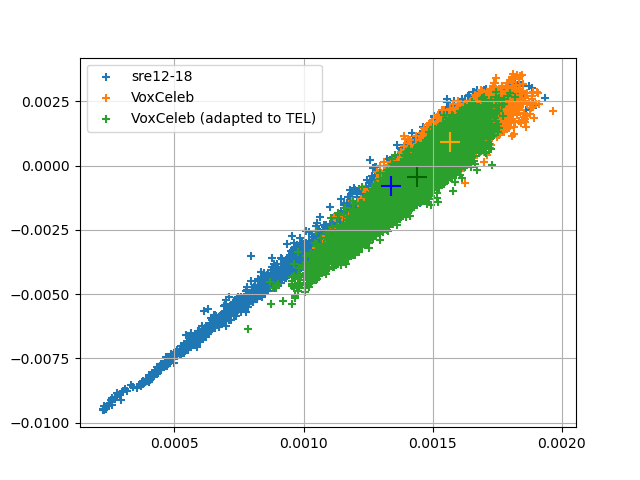}
    \caption{t-SNE visualization of embeddings for telephone (SRE12-18), microphone (VoxCeleb2), and microphone adapted to telephone. Respective centroids are denoted by big ``+''.}
    \label{fig:tsne}
\end{figure}

In Table~\ref{tab:baseline}, we demonstrate the benefit of adapting microphone data to the telephone domain using vanilla CycleGAN.
Results are reported in the format: \emph{EER / minDCF}.
Improvement is significant in EER as well as minDCF for all three test sets.
As a motivational illustration, Fig.~\ref{fig:tsne} shows the working of adaptation using 2-D \ac{t-SNE}~\cite{van2008visualizing} visualization.
We can observe that the centroid of microphone embeddings has shifted considerably close to the telephone centroid.
We perform various runs of t-SNE to find that this structure is highly robust to the choice of perplexity value (5-50) and metric type (Euclidean/Cosine).
We produce the illustration for perplexity value of 30 and Euclidean metric.
To justify the significance of the relative shift in centroid locations, we preserve global structure in addition to local via \ac{PCA} initialization in open t-SNE~\cite{Policar731877}.
Additionally, we tried \ac{UMAP} visualization~\cite{mcinnes2018umap} to arrive at identical observations.

\begin{table}
   \caption{Comparison of adaptation with vanilla CycleGAN vs no adaptation when using microphone data (VoxCeleb) for x-vector training and no data augmentation. Bold typeface denotes the best metric value obtained.}
  \label{tab:baseline}
  \centering
  \resizebox{0.47\textwidth}{!}{
\begin{tabular}{|c|c|c|c|}
\hline
  \begin{tabular}[c]{@{}c@{}}Adaptation $\downarrow$\end{tabular}               & SRE19-eval & SRE16-yue & SRE16-tgl \\
\hline
None        &      7.33 / 0.357      &     6.79 / 0.394      &     14.81 / 0.669      \\
\hline
\emph{CycleGAN}         &      \textbf{7.02} / \textbf{0.351}      &     \textbf{6.33} / \textbf{0.354}      &     \textbf{13.79} / \textbf{0.634}      \\
\hline
\end{tabular}
}
\end{table}

\begin{table}
  \caption{Effect of modifying cycle-consistency loss, identity loss, and adversarial loss independently as well as in combinations on top of baseline adaptation system of Sec.~\ref{sec:baseline}.}
  \label{tab:dfcyclegan}
  \centering
  \resizebox{0.47\textwidth}{!}{
\begin{tabular}{|c|c|c|c|}
\hline
  \begin{tabular}[c]{@{}c@{}}Adaptation $\downarrow$\end{tabular}               & SRE19-eval & SRE16-yue & SRE16-tgl \\
\hline
\emph{CycleGAN} & 7.02 / 0.351 & 6.33 / 0.354 & 13.79 / 0.634      \\
\hhline{|=|=|=|=|}
\emph{DF-Cycle}         &      6.71 / 0.344      &     6.46 / 0.357      &     13.40 / 0.614      \\
\hline
\emph{DF-Identity}         &      6.73 / 0.336      &     6.39 / 0.344      &     14.66 / 0.644      \\
\hline
\emph{DF-Disc.~(2D)}         &      7.33 / 0.368 & 6.49 / 0.370 & 14.33 / 0.646
      \\
\hline
\emph{DF-Disc.~(3D)}         &      6.92 / 0.341      &     5.87 / \textbf{0.340}      &     \textbf{13.18} / 0.622      \\
\hline
\begin{tabular}[c]{@{}c@{}}\emph{DF-Identity}\\+ \emph{DF-Cycle}\end{tabular} & 6.99 / 0.342 & 6.18 / 0.368 & 13.66 / 0.622 \\
\hline
\emph{\textbf{DF-CycleGAN}}         &      \textbf{6.70} / \textbf{0.332}      &     \textbf{5.81} / 0.351      &     13.68 / \textbf{0.612}      \\
\hline
\end{tabular}
}
\end{table}

\begin{table*}[t]
    \caption{Studying the effect of pooling together telephone and (adapted) microphone data for the training of the x-vector network. $p_{\text{adapt}}$ is the probability of adaptation of microphone samples seen during SGD based training of x-vector network. Results are provided in format (EER (in \%) / minDCF) for various values of $p_{\text{adapt}}$ in the presence or absence of telephone codecs and data augmentation.}
    \centering
    \resizebox{0.94\textwidth}{!}{
    \begin{tabular}{|c|c|c|c|c|c|c|c|}
        \hline
              x-vector training data type $\downarrow$& $p_{\text{adapt}}$ & Tel codecs & $\lambda_{\text{id}}$ & Data Augmentation & SRE19-eval & SRE16-yue & SRE16-tgl \\
        \hline
        \emph{Microphone} & 0 & \cmark & - & \xmark & 7.33 / \textbf{0.357} & 6.79 / 0.394 & 14.81 / \textbf{0.669} \\
        \hline
        \emph{Telephone} & - & - & - & \xmark & \textbf{6.85} / 0.381 & \textbf{5.9} / \textbf{0.348} & \textbf{14.48} / 0.699 \\
        \hhline{|=|=|=|=|=|=|=|=|}
        \emph{Telephone} + \emph{Microphone} & 0 & \cmark & - & \xmark & 5.64 / 0.311 & 4.95 / 0.284 & 13.09 / 0.624 \\
        \hline
        \emph{Telephone} + \emph{Microphone} & 1 & \xmark & 5 & \xmark & 5.52 / 0.304 & 4.40 / 0.265 & 12.60 / 0.615 \\
        \hline
        \emph{Telephone} + \emph{Microphone} & 1 & \cmark & 5 & \xmark & 5.79 / 0.315 & 5.14 / 0.291 & 12.12 / 0.608 \\
        \hline
        \emph{Telephone} + \emph{Microphone} & 0.2 & \cmark & 5 & \xmark & 5.40 / 0.294 & 4.41 / 0.259 & 12.01 / 0.598
 \\
        \hline
        \emph{Telephone} + \emph{Microphone} & 0.5 & \cmark & 5 & \xmark & \textbf{5.26} / \textbf{0.288} & \textbf{4.37} / 0.264 & 11.83 / 0.587 \\
        \hline
        \emph{Telephone} + \emph{Microphone} & 0.8 & \cmark & 5 & \xmark & 5.34 / 0.299 & 4.79 / \textbf{0.257} & \textbf{11.79} / \textbf{0.584} \\
        \hhline{|=|=|=|=|=|=|=|=|}
        \emph{Telephone} + \emph{Microphone} & 0 & \xmark & - & \cmark & 4.63 / 0.261 & 4.10 / 0.271 & 10.41 / 0.530 \\
        \hline
        \emph{Telephone} + \emph{Microphone} & 0 & \cmark & - & \cmark & 4.65 / 0.261 & 4.44 / 0.276 & 10.72 / 0.539 \\
        \hline
        \emph{Telephone} + \emph{Microphone} & 1 & \xmark & 10 & \cmark & 4.74 / 0.264 & \textbf{4.08} / 0.254 & 9.81 / 0.522 \\
        \hline
        \emph{Telephone} + \emph{Microphone} & 1 & \cmark & 10 & \cmark & 4.98 / 0.282 & 4.80 / 0.273 & 9.50 / \textbf{0.497} \\
        \hline
        \emph{Telephone} + \emph{Microphone} & 0.5 & \xmark & 10 & \cmark & 4.70 / 0.264 & 4.27 / 0.277 & 10.17 / 0.536 \\
        \hline
        \emph{Telephone} + \emph{Microphone} & 0.5 & \cmark & 5 & \cmark & 4.92 / 0.275 & 4.37 / 0.262 & 9.69 / 0.505\\
        \hline
        \emph{Telephone} + \emph{Microphone} & 0.5 & \cmark & 10 & \cmark & \textbf{4.60} / \textbf{0.250} & 4.22 / \textbf{0.252} & \textbf{9.34} / 0.499 \\
        \hline
        \end{tabular}
    }
    \label{tab:multidata}
\end{table*}

\subsection{Deep feature-based CycleGAN adaptation}
\label{sec:dfcgresults}
To improve CycleGANs, we introduced three modifications (Eq.~\ref{eq:dfcg}) to the original formulation.
In Table~\ref{tab:dfcyclegan}, we study the effect of those modifications independently as well as combined.
We make three observations.
One, modifying cycle-consistency, identity, and adversarial loss independently improves results on all six metrics with certain exceptions.
Two, using 3D convolutions in discriminator (\emph{DF-Disc.~(3D)}) is vastly superior to 2D convolutions due to the preservation of 5D data structure (Sec.~\ref{sec:dfcyclegan}).
Three, the adaptation model with all three modifications (\emph{DF-CycleGAN}) is the best while the one with only modified identity and cycle-consistency loss (\emph{DF-Identity} + \emph{DF-Cycle}) suggests the sensitivity to the choice of hyper-parameters like $\lambda_{\text{id}}$, $\lambda_{\text{cyc}}$.
It is important to note that modification of loss functions changes their dynamic range and, to obtain best results, re-tuning of weights $(\lambda_{\text{cyc}},\lambda_{\text{id}})$ is required, which is an expensive procedure for CycleGANs.
We do not pursue an extensive weight search and hence, the results can be potentially improved.

In \emph{deep feature} based CycleGANs, the computational complexity and memory requirements increases due to the usage of auxiliary networks ($\mathcal{A}_T$, $\mathcal{A}_M$), 3D convolutions, and increased dimensionality of input features to generators and discriminators.
As a trade-off between complexity and performance, we look for minimal modifications to CycleGAN and, thus use \emph{DF-Disc.~(3D)} for further detailed experiments.

\subsection{Pooling telephone and adapted microphone data with data augmentation and telephone codecs}
In Sec.~\ref{sec:baseline} and Sec.~\ref{sec:dfcgresults}, we demonstrated the benefit of baseline and proposed CycleGAN adaptation of microphone data for the purpose of x-vector training.
It is imperative to study the pooling together of this adapted data with available telephone training data.
Moreover, in practice, for best performance, it is standard to apply noise augmentation on all data and treat microphone speech (after downsampling) with telephone codecs like GSM, AMR-NB, etc.
These operations are done in the time-domain and \ac{LMFB} features are extracted subsequently.

We tabulate the corresponding results in Table~\ref{tab:multidata}.
We first demonstrate that our choice of the telephone (\emph{SRE-Mix-SWBD}) and the microphone data (\emph{VoxCeleb2}) gives a good performance, and therefore, can be combined for potentially further gains.
When pooling together telephone and (adapted) microphone data, we first experiment without data augmentation.
A parameter $p_\text{adapt}$ is introduced which refers to the probability of adapting microphone features on the fly when training the x-vector network with \ac{SGD}.
We introduce this parameter due to the observation that, in the case of no augmentation, adaptation does not bring consistent improvements.
Comparing results with and without codecs, we observe that the role of codecs varies with test sets also.
For instance, with adaptation and codecs, EER on \emph{SRE16-tgl} is improved from 13.09 \% to 12.12 \% but results are worse on other test sets.
Using codecs along with a convenient value of $p_\text{adapt}=0.5$ alleviates these issues and we obtain a decent performance overall.

For the systems with augmentation, we find the role of codecs to be inconsistent again.
We can note from the rows with $p_{\text{adapt}}=1$ that adaptation helps regardless of other factors.
Parameter $p_\text{adapt}=0.5$ further boosts the performance possibly due to the increased diversity in training samples.
Recall that the CycleGAN is trained with \ac{LMFB} features without augmentation and codecs.
Using such a system for speech with augmentation and/or codecs requires the adaptation mapping to be robust to these operations.
A simple trick to achieve this is to increase the value of identity loss as it makes the generators more conservative~\ref{sec:cyclegan}.
The significance of the higher value of $\lambda_{\text{id}}$ can be noted from the last two rows.
Our overall best system, thus, requires data augmentation, telephone codecs, stochastic adaptation ($p_\text{adapt}=0.5$), and a higher identity loss ($\lambda_{\text{id}}=10$).

\section{Conclusion}
We study if reducing the \emph{domain shift} of microphone data w.r.t. telephone domain is beneficial for the x-vector training for Telephony Speaker Verification.
Using a generative unpaired translation technique called CycleGAN, we observe relative improvements of 5-10\% in EER, minDCF on three challenging language-mismatched test sets.
Then, we propose three modifications to the formulation of CycleGAN based on \emph{deep features}.
We find this improves cycle-consistency constraint and preserves speaker identity during adaptation.
Finally, we study a challenging scenario where we pool adapted microphone data with telephone data and gain insights into the interplay of various hyper-parameters $(p_{\text{adapt}}, \lambda_{\text{id}})$, codecs, and augmentation.
Our approach demonstrates great potential for \emph{task-specific} domain adaptation.
In the future, we can explore the adaptation of non-English training data, learn many-to-many CycleGAN mappings, and explore the latest telephone codecs~\cite{sivaraman2020speech}.

\clearpage
\bibliographystyle{IEEEtran}
\bibliography{template}

\end{document}